\documentclass[11pt,a4paper]{article}
\usepackage{amsmath,amssymb,amsfonts,bm}
\usepackage{graphicx}
\usepackage[top=1cm,body={16cm,23cm}]{geometry}

\begin{document}

\title{A HYBRID MODEL FOR $\gamma^* p$ SCATTERING \\
AT SMALL BJORKEN-x
}

\author{A. SZCZUREK \\
Institute of Nuclear Physics \\
PL-31-342 Cracow, Poland,\\
University of Rzesz\'ow,\\
PL-35-959 Rzesz\'ow, Poland,\\
E-mail: antoni.szczurek@ifj.edu.pl }

\maketitle

\begin{abstract}
\noindent
We extend the dipole model for virtual photon - proton
scattering to include the resolved photon component explicitly.
The parameters of the resolved photon component are taken from
the literature, while the parameters
of the dipole-nucleon interaction are fitted to the HERA data.
A good agreement with experimental data is obtained beyond
the region of the fit.
\end{abstract}

\section{Introduction}

The recent decade of investigating deep inelastic scattering
at small Bjorken $x$ at HERA has provided precise data for
the $F_2$ structure function or equivalently for
$\sigma^{\gamma^* p}_{tot}$ at large center-of-mass energies.
Many phenomenological analyses have been performed in order
to fit the data.
One group of models tries to fit the data
using the so-called dipole representation. In this approach,
initiated in Ref.\cite{NZ90}, one fits parameters
of the dipole-nucleon interaction \cite{GBW,FKS99,KD00} as a function
of the transverse $q \bar q$ distance.

The fits in the dipole representation take into account only
a simple quark-antiquark Fock component of the photon. However,
the higher Fock components seem to be important to understand
the diffraction \cite{GBW_glue}.
The first theoretical step in going beyond the $q \bar q$ component
has been undertaken only recently \cite{Bartels}.
However, no quantitative estimates exist up to now.  
On the phenomenological side, the jet production in virtual-photon-proton
scattering, especially at small photon virtuality, shows clearly
the presence of the resolved photon component
(see e.g. \cite{jets_resolved_photon}).
The resolved photon component seems also crucial for understanding
the world data for the $F_2^p(x,Q^2) - F_2^n(x,Q^2)$ \cite{SU_F2p_F2n}.
All these arguments put into question the simple fits to
the total photon-nucleon cross section with the colour dipole
component alone, and call for a multi-component parametrization.

\section{Formulation of the model}

It is known that the LO total $\gamma^* N$ cross section
in the so-called dipole or mixed representation can be written in the form
\begin{equation}
\sigma_{tot}^{\gamma^* N} = \sum_q \int dz \int d^2 \rho
\; \sigma_{T,L} \; | \Psi_{\gamma^* \rightarrow q \bar q}^{T,L}(Q,z,\rho) |^2
\cdot \sigma_{(q \bar q) N}(x,\rho) \; .
\label{dipole_nucleon}
\end{equation}
In this paper we take the so-called quark-antiquark
photon wave function of perturbative form \cite{NZ90}.
As usual, in order to correct the photon wave function for large
dipole sizes we introduce an effective (anti)quark mass ($m_{eff} = m_0$).

The dipole representation (\ref{dipole_nucleon}) has been used in
recent years to fit the virtual photon - nucleon total cross section
\cite{GBW,FKS99}. The best fit has been achieved in the saturation
model of Golec-Biernat-W\"usthoff \cite{GBW}.
In their approach the dipole-nucleon cross section was parametrized as
\begin{equation}
\sigma(x,\rho) = \sigma_0
\left[ 1 - \exp\left( - \frac{\rho^2}{4 R_0^2(x)} \right) \right] \; ,
\label{GBW_dipole_nucleon}
\end{equation}
where the Bjorken $x$ dependent radius $R_0$ is given by
$R_0(x) = \frac{1}{1 GeV} \left( \frac{x}{x_0} \right)^{\lambda/2}$.
Model parameters (normalization constant $\sigma_0$
and parameters $x_0$ and $\lambda$) have been determined by the fit
to the inclusive data on $F_2$ for $x <$ 0.01 \cite{GBW}.

In the GBW approach, the dipole-nucleon cross section is parametrized
as a function of Bjorken $x$. As discussed in \cite{Sz02},
it would be useful to have rather a parametrization in the gluon
longitudinal momentum fraction $x_g \ne x$ instead of the Bjorken $x$.
Having $x_g$ instead of Bjorken-x better reflects the kinematics
of the process and is consistent with the standard approach to
photon-gluon fusion.
This involves the following replacement in Eq.(\ref{GBW_dipole_nucleon})
$\sigma(x,\rho) \rightarrow \sigma(x_g,\rho)$
which means also a replacement of  $x$ by $x_g$ in $R_0$.
As discussed in \cite{Sz02}, an exact calculation of $x_g$ in the dipole
representation is, however, not possible, and we approximate 
$x_g \rightarrow (M_{qq}^2 + Q^2)/(W^2 + Q^2)$, where $M_{qq}^2 =
m_q^2/(z(1-z))$.

We intend to construct a simple two-component model. One component
of our phenomenological model is the dipole $q \bar q$ component, 
while the other one is meant to represent the nonperturbative resolved
photon component explicitly. 
Trying to keep our model as simple as possible 
we have tried to check if the standard vector dominance model (VDM)
contribution can be a reasonable representation of
the resolved photon. 
The cross section for the VDM component is calculated in the standard way
\begin{equation}
\sigma^{VDM}_{\gamma^* N}(W,Q^2) = \sum_V
\frac{4 \pi}{\gamma_V^2} \frac{ M_V^4 \sigma^{VN}_{tot}(W) }{(Q^2 +
  M_V^2)^2}
\cdot (1-x) \; .
\label{VDM_component}
\end{equation}
We take the simplest diagonal version of VDM with $\rho$, $\omega$ and
$\phi$ mesons included. Other details can be found in \cite{PS03}.

\section{Fit to the HERA data}

In the previous section we presented formulae for the virtual
photon - nucleon cross section.
We transform the structure function data
from \cite{HERA_data} to $\sigma_{tot}^{\gamma^* N}$.
Then we perform two independent fits to the HERA data. In fit 1, only
dipole nucleon interaction is included (see Eq.(\ref{dipole_nucleon}))
\begin{equation}
FIT1: \;\; \sigma_{tot}^{\gamma^*N} = \sigma_{dip}^{\gamma^*N} \; .
\label{fit1}
\end{equation}
In fit 2 in addition we include the resolved photon component
in the spirit of the vector meson dominance model
(see Eq.(\ref{VDM_component}))
\begin{equation}
FIT2: \;\; \sigma_{tot}^{\gamma^*N} = \sigma_{dip}^{\gamma^*N}
                               + \sigma_{VDM}^{\gamma^*N} 
 \; .
\label{fit2}
\end{equation}

In these fits we limit to 0.15 GeV$^{2}$ $ < Q^2 < $ 10 GeV$^2$.
The upper limit is dictated by the simplicity of our model.
The maximal Bjorken $x$ in the data sample included in our fit is 0.021,
and minimal W = 17.4 GeV.

The region of small $Q^2$ is sensitive to the value of the effective
quark mass. 
A good quality fit can be obtained in the broad range
of $m_0$. 
The value of $\chi^2$ in fit 2 (dipole+VDM) is smaller
than that in fit 1 (dipole only).
This can be taken as the evidence for resolved
photon component.


\begin{figure}[thb]
  \begin{center}
    \includegraphics[width=8cm]{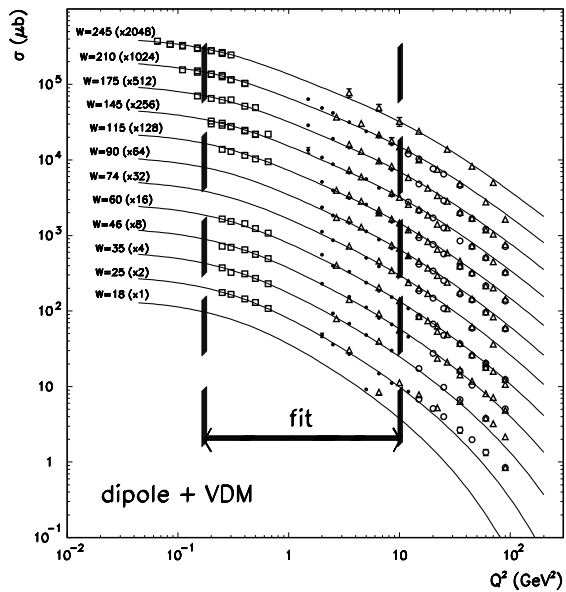}
    \caption{Quality of fit 2 ($q \bar q $ dipole and VDM) --- 
      cross sections as a~function of Q$^2$. The HERA data taken from
      \cite{HERA_data}.}
  \end{center}
\end{figure}


The quality of the fit can be judged by inspecting
Fig.1.
Since there is a rather weak dependence of the cross section on $W$,
therefore in the figure both theoretical
curves and experimental points are rescaled by an extra factor 2$^n$.
Careful comparison of fit 1 and fit 2 shows the superiority of the fit
2 in the region of small $Q^2$ and large energies \cite{PS03}.
In presenting the results we have made an arbitrary choice
of $m_0$. The results for other sets of parameters (different $m_0$)
are almost indistinguishable in the range of the fit. They differ
somewhat, however, outside the range of the fit where no experimental data
are available.
The theoretical curves with dipole component only underestimate
somewhat the low $Q^2$ data. 

Having shown that a good-quality two-component fit to the HERA data
with very small number of parameters is possible, we wish to show 
a decomposition of the cross section into the two model components.
In Fig.2, as an example, we show separate contributions of both
components as a function of $W$.
While at low energy the VDM contribution dominates
due to the subleading reggeon exchange, at higher energies they are of
comparable size. The VDM contribution
dominates at small values of photon virtualities, while
at larger $Q^2$ the dipole component becomes dominant.


\begin{figure}[thb]
  \begin{center}
    \includegraphics[width=8cm]{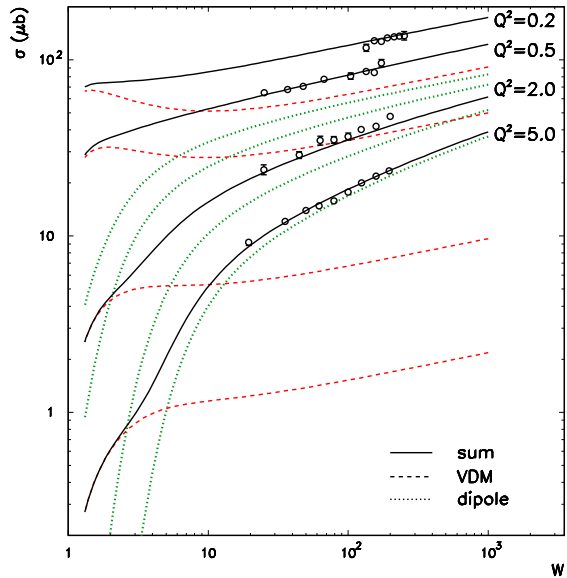}
    \caption{Decomposition of total $\gamma^*{}p$ cross section into
      dipole (dotted) and VDM (dashed) contributions for 4 different
      values of photon virtuality in GeV$^2$.}
  \end{center}
\end{figure}


Up to now we have concentrated on very low-$x$ region relevant for DIS
at HERA.
It is interesting to check what happens if we go to somewhat larger
Bjorken $x >$ 0.05.
In this region one cannot neglect the valence
quark contribution to the cross section.
Then the cross section is a sum of three components:
\begin{equation}
\sigma_{tot}^{\gamma^* N}(W,Q^2) =
\sigma_{dip}^{\gamma^* N}(W,Q^2) +
\sigma_{VDM}^{\gamma^* N}(W,Q^2) +
\sigma_{val}^{\gamma^* N}(W,Q^2) \; .
\label{dip_VDM_val}
\end{equation} 
We have found a good quality description of the fixed target data
\cite{PS03}. 

\section{Conclusions}

Recent fits to the total $\gamma^* p$ cross section in the literature
include only the $q \bar q$ component in the Fock decomposition
of the photon wave function. 
It is known from the phenomenology of the inclusive and exclusive
reactions that the vector dominance model in many cases gives
a good estimate of the effects characteristic for resolved photon.
We have analyzed if a two-component model,
which includes the $q \bar q$ component and the more complicated
components replaced by the standard VDM, can provide
a good description of the HERA data for $\gamma^* p$ scattering.

In order to quantify the effect of the resolved photon we have performed
two different fits to the HERA data. In fit No.1 we include only the dipole
component.  We have parametrized the dipole-nucleon cross section in terms of
a variable which is closer to the gluon longitudinal momentum fraction
$x_g$ than to the Bjorken $x$.
In fit No.2, in addition we include the VDM component
while keeping the same functional form of parametrization for
the dipole-nucleon interaction.
At small $Q^2$ and large energies a better fit is obtained
if the resolved photon component is included.
When going to slightly larger Bjorken
$x >$ 0.05, the model must be supplemented for valence quark
contribution. If this is done,
the model describes also the fixed target data quite well.


\end{document}